\documentclass[prb,floatfix,showpacs,amsmath,twocolumn]{revtex4}

\usepackage{amssymb}
\usepackage[mathscr]{eucal}
\usepackage[dvips]{graphicx}

\newcommand{\expect}[1]{\langle #1 \rangle}

\newcommand{\Sprod}[2]{\mathbf{s}_{#1} \cdot \mathbf{s}_{#2}}

\begin{document}

\title{Exploring frustrated spin--systems using Projected Entangled Pair States (PEPS)}

\author{V. Murg$^1$, F. Verstraete$^2$, J. I. Cirac$^1$}
\affiliation{$^1$Max-Planck-Institut f\"ur Quantenoptik,
Hans-Kopfermann-Str. 1, Garching, D-85748, Germany\\
$^2$Fakult\"at f\"ur Physik, Universit\"at Wien, Boltzmanngasse 3,
A-1090 Wien}
\date{\today}

\begin{abstract}
We study the nature of the ground state of the frustrated $J_1-J_2$ model and the $J_1-J_3$ model using a variational algorithm based on projected entangled-pair states (PEPS). By investigating spin--spin correlation functions, we observe a separation in regions with long--range and short--range order.
A direct comparison with exact diagonalizations in the subspace of short--range valence bond singlets reveals that the system is well described by states within this subset in the short--range order regions. We discuss the question whether the system forms a spin--liquid, a plaquette valence bond crystal or a columnar dimer crystal in these regions.
\end{abstract}

\maketitle


\section{Introduction}

Frustrated spin--systems have attracted a lot of interest in the last years, because they may possess exotic ground states that are very different from conventional N\'eel--ordered states. Such states are especially intriguing, as connections to high-$T_c$ superconductivity have been put forward~\cite{anderson87}. They are usually characterized by a break down of long--range order: the system reorganizes in a quantum state where only local antiferromagnetic correlations are present. The class of such states, named Short Range Valence Bond States (SRVB), encompasses a broad range of phases: they range from valence bond crystals (VBC) with broken translational symmetry to pure spin liquids that have all symmetries restored.

Studies of frustrated systems are especially challenging, because Quantum Monte--Carlo (QMC) studies are hindered by the sign-problem~\cite{troyer04} and Density Matrix Renormalization Group (DMRG)~\cite{white92,white92b,schollwoeck04} investigations are hard to perform for systems with dimensionality larger than one. Other methods that have been developed to take on these systems are, for example, the coupled cluster method~\cite{richter07}, DMRG combined with QMC~\cite{jongh00} and exact diagonalizations within the subspace of SRVB~\cite{mambrini06,mambrini00}.
In this paper, we would like to give the recently developed PEPS--algorithm~\cite{verstraetecirac04,murgverstraete07} a try. This algorithm has already been successfully applied to the Heisenberg antiferromagnet~\cite{verstraetecirac04}, the Bose--Hubbard model~\cite{murgverstraete07} and the frustrated Shastry--Sutherland model~\cite{isacsson06}.

In the following, we focus on the $J_1-J_2-J_3$ model. We discuss our observations and possible implications. Before that, we give a quick review of the PEPS algorithm in section~\ref{sec:PEPS}. For a thorough explaination of the algorithm, we refer to Refs.~\onlinecite{verstraetecirac04,murgverstraete07,verstraeteciracmurg08}.
With the implementation of the algorithm, we follow in
large part Ref.~\onlinecite{murgverstraete07}.


\begin{figure}
    \begin{center}
        \includegraphics[width=0.44\textwidth]{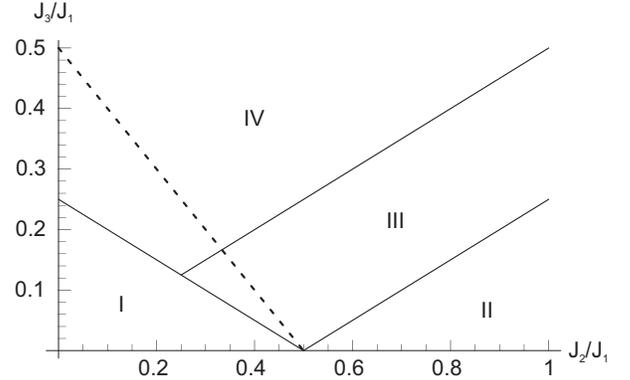}
    \end{center}
    \caption{
        Classical phase diagram of the $J_1-J_2-J_3$ model consiting of (I) a
N\'eel phase, (II) a phase consisting of two independent sublattices, (III) a
helical phase with order $(q,q)$ and (IV) a helical phase with order $(q,\pi)$.
The dashed line denotes the boundary of the N\'eel phase when quantum
fluctuations are taken into account.
        }
    \label{fig:J1J2J3}
\end{figure}

\section{The PEPS-algorithm revisited} \label{sec:PEPS}

The PEPS algorithm is a variational algorithm within the class of \emph{Projected Engangled Pair States} (PEPS)~\cite{verstraetecirac04,murgverstraete07}. These states are expedient for a numerical study of many--particle systems: on the one hand, they fulfill the area--law that is fundamental to non--critical systems. On the other hand, they possess a number of parameters that scales polynomially with the system size. This feature makes numerical simulations feasible.

The structure of PEPS is inspired from the structure of \emph{Matrix Product States} (MPS)~\cite{oestlund95,dukelsky98} that form the basis for the successful DMRG algorithm~\cite{white92,white92b,schollwoeck04}: to each lattice site, a tensor is associated that possesses a physical index and a certain number of virtual indices. The virtual indices are contracted according to a scheme that mimics the underlying lattice--structure. The dimension of the physical index equals the physical dimension of the particle residing on that lattice site. The dimension of the virtual indices, $D$, is the internal refinement parameter of PEPS: $D = 1$ specializes the PEPS to a product state; the choice $D = d^N$ (with $N$ being the total number of lattice sites and $d$ the physical dimension of one particle) enlarges the space of PEPS to the complete Hilbert--space of the system.

The main idea of the algorithm is to optimize the tensors such that the energy tends to a minimum. This can be done by sequentially optimizing the tensors or by cooling the system with the simulation of an imaginary time evolution. Here, we follow the latter. Time and memory required for this method turn out to be polynomial both in $N$ and $D$:  time scales as $N^2 D^{10}$ and memory as $N D^{8}$.

This polynomal scaling allows us to investigate systems of sizes that are out of reach with other algorithms. We discuss our observations in the following sections.


\section{$J_1-J_2-J_3$ Model} \label{sec:J1J2J3}

In the $J_1-J_2-J_3$ model on a square lattice, frustration is caused by the competition between first, second and third neighbor interactions of magnitudes~$J_1$, $J_2$ and $J_3$ respectively:

\begin{displaymath}
H = J_1 \sum_{\langle ij \rangle} \mathbf{s}_i \cdot \mathbf{s}_j +
J_2 \sum_{\langle \langle ij \rangle \rangle} \mathbf{s}_i \cdot \mathbf{s}_j +
J_3 \sum_{\langle \langle \langle ij \rangle \rangle \rangle} \mathbf{s}_i \cdot \mathbf{s}_j
\end{displaymath}

The phase diagram of this model is involved and still controversal. Of special interest are the regimes of maximal frustration that are suspected of having non-classical ground states.

Let us first review the classical
limit~\cite{ferrer93,chubukov91,moreo90,gelfand89} ($S \to \infty$). In this limit, the system possesses four phases, as shown in figure~\ref{fig:J1J2J3}: the usual N\'eel phase, two spiral antiferromagnetic phases ordered at $(q,q)$ and $(q,\pi)$ and a phase consisting of two independent sublattices. The N\'eel phase is bounded by the \emph{classical critical line} $(J_2+2J_3)/J_1=1/2$.

When quantum fluctuations are taken into account, the phase diagram changes considerably~\cite{ferrer93,figueirido89,read91,mambrini06,sachdev91}: the N\'eel phase substantially extends to larger values of $J_3$, up to the line of maximal frustration $(J_2+J_3)/J_1=1/2$. In the vicinity of this line, it is believed that the classical ordered ground state is destabilized and a singlet ground state is formed~\cite{mambrini06}. The precise nature of this state is still controversal: suggestions include columnar valence bond crystals~\cite{leung96}, plaquette states~\cite{mambrini06} and spin liquids~\cite{capriotti04,zhong93,capriottisachdev04,chandra88,locher90}.
Special attention has been devoted to the end-points of the line: the point at $J_3/J_1=1/2$ that separates the N\'eel from the spiral antiferromagnetic phase~\cite{capriotti04,capriottisachdev04,locher90}, and the tri-critial point at $J_2/J_1=1/2$ at which $3$ phases
meet~\cite{zhong93,jongh00,chandra88,chandra90}.

\begin{figure}
    \begin{center}
        \includegraphics[width=0.44\textwidth]{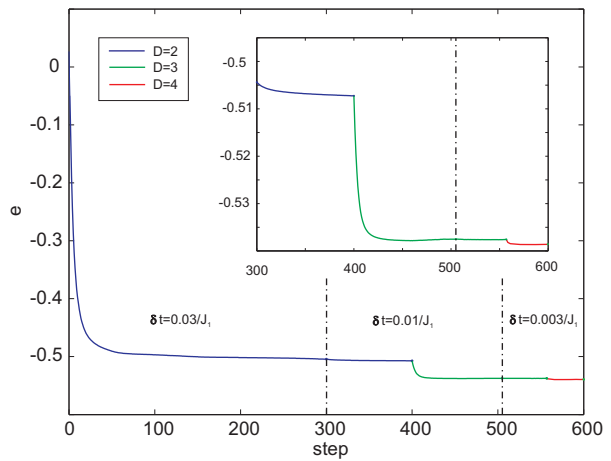}
    \end{center}
    \caption{
        Energy per site as a function of the imaginary time evolution steps. The lattice size is $10 \times 10$ and $J_3/J_1=0.5$. The inset shows the magnified tail of the energy function. The vertical lines separate the plot into three regions: in these regions, the Trotter-step $\delta t$ is chosen equal to $0.03/J_1$, $0.01/J_1$ and $0.003/J_1$ respectively.
        }
    \label{fig:J1J3:energy}
\end{figure}

In the following, we focus on two lines in the phase diagram that include these points: $J_2=0$ ($J_1-J_3$ model) and $J_3=0$ ($J_1-J_2$ model). In both cases, we apply the PEPS-algorithm and discuss our observations. Our investigations are thereby inspired by the discussions in Ref.~\onlinecite{mambrini06}. We obtain the PEPS--approximation of the ground state via imaginary time evolution, as described previously. In the course of the time evolution, we increase the virtual dimension~$D$, starting from $D=2$, and decrease the Trotter-step~$\delta t$, starting from $\delta t=0.03/J_1$, until the relative changes in the ground state energy are of order $10^{-3}$. This is usually achieved for $D=4$ and $\delta t = 0.003/J_1$. Typically, $D=4$ only leads to a minor reduction of the energy. The energy as a function of the imaginary time evolution steps is plotted in figure~\ref{fig:J1J3:energy} for the special case of a $10 \times 10$ lattice and $J_3/J_1=0.5$.


\subsection{Spin--Spin Correlations} \label{sec:corr}

In order to get a first idea of the nature of the ground state, we calculate the spin--spin correlation functions $\expect{ \mathbf{s}_k \cdot \mathbf{s}_l }$ and the corresponding static structure factor,
\begin{equation} \label{eqn:structurefactor}
S(\mathbf{q}) = \frac{1}{N^2} \sum_{k l} e^{i \mathbf{q} \cdot ( \mathbf{r}_k - \mathbf{r}_l ) } \expect{ \mathbf{s}_k \cdot \mathbf{s}_l },
\end{equation}
in the PEPS--approximation of the ground state for $0 \leq J_3/J_1 \leq 1$ at $J_2=0$ and $0 \leq J_2/J_1 \leq 1$ at $J_3=0$.

\subsubsection*{$J_1-J_3$ Model}

\begin{figure}
    \begin{center}
        \includegraphics[width=0.44\textwidth]{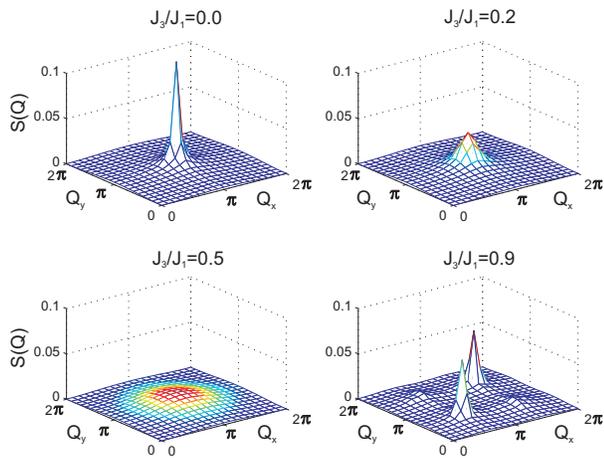}
    \end{center}
    \caption{
        Structure factor $S(\mathbf{q})$ for $J_3/J_1=0$, $0.2$, $0.5$ and $0.9$. The results were obtained for a $14 \times 14$--lattice and virtual dimension $D=3$.
        }
    \label{fig:J1J3:SQ}
\end{figure}

In the case of $J_2=0$, we observe that $S(\mathbf{q})$ is peaked at $\mathbf{q} = (\pi,\pi)$ for $J_3/J_1 < 0.5$, indicating long--range N\'eel order. This order disappears at $J_3/J_1 \sim 0.5$ at which the structure factor becomes smooth. At around $J_3/J_1 \sim 0.7$ peaks at $(\pm \pi/2, \pm \pi/2)$ reappear, indicating a revival of incommensurate long--range order.
Figure~\ref{fig:J1J3:SQ} depicts the functional characteristics of the structure factor $S(\mathbf{q})$ at selected points $J_3/J_1$, calculated with a $D=3$ PEPS on a $14 \times 14$ lattice.

\begin{figure}
    \vspace{0.8cm}
    \begin{center}
        \includegraphics[width=0.44\textwidth]{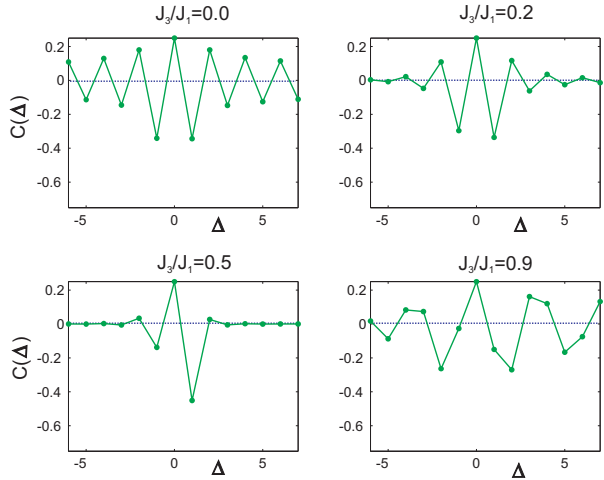}
    \end{center}
    \caption{
        Spin--spin correlations $\expect{\mathbf{s}_i \cdot \mathbf{s}_j}$ as functions of the distance~$\Delta=|i-j|$ for the $14 \times 14$ $J_1-J_3$ model at $J_3/J_1=0$, $0.2$, $0.5$ and $0.9$. The results have been obtained using a $D=3$--PEPS Ansatz.
        }
    \label{fig:J1J3:corr}
\end{figure}

The collapse of long--range order is confirmed by a direct observation of the spin--spin correlations. These are shown in figure~\ref{fig:J1J3:corr}. Here, correlations with the central spin are plotted as functions of the distance. As can be seen, the spins are antiferromagnetically ordered for $J_3/J_1 < 0.5$. For $J_3/J_1 > 0.5$, every second spin possesses antiferromagnetic order. However, the long--range order of the spins disappears in the vicinity of $J_3/J_1 \sim 0.5$.

\begin{figure}
    \begin{center}
        \includegraphics[width=0.44\textwidth]{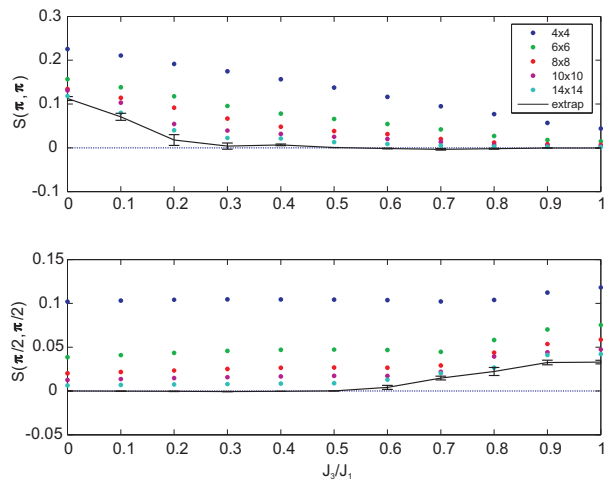}
    \end{center}
    \caption{
		Structure factors $S(\mathbf{\pi,\pi})$ (upper plot) and $S(\mathbf{\pi/2,\pi/2})$ (lower plot) as functions of $J_3/J_1$ for various particle--numbers~$N$. The solid lines represent extrapolations to the thermodynamic limit (see text).
        }
    \label{fig:J1J3:SQ_pi_pi2}
\end{figure}

\begin{figure}
    \begin{center}
        \includegraphics[width=0.44\textwidth]{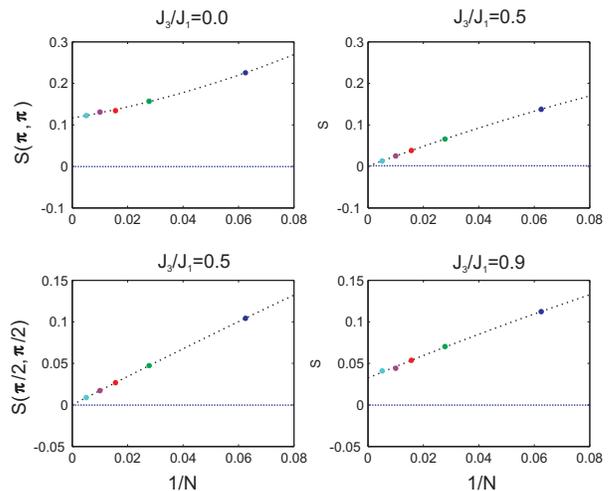}
    \end{center}
    \caption{
		Structure factors $S(\mathbf{\pi,\pi})$ (upper plots) and $S(\mathbf{\pi/2,\pi/2})$ (lower plots) as functions of $1/N$ for selected points $J_3/J_1$. The dotted lines represent the best fits by polynomials of degree~$3$ in $1/N$.
        }
    \label{fig:J1J3:SQ_Nscaling}
\end{figure}

The separation in long--range and short--range order regimes gets more and more evident with growing particle--number. This can be gathered from figure~\ref{fig:J1J3:SQ_pi_pi2}. Here, $S(\pi,\pi)$ and $S(\pi/2,\pi/2)$ are plotted as functions of $J_3/J_1$, evaluated for various particle--numbers~$N$. An extrapolation to the thermodynamic limit ($N \to \infty$) shows that $S(\pi,\pi)$ remains finite within the region $J_3/J_1 \lesssim 0.3$ and is zero otherwise. On the other hand, $S(\pi/2,\pi/2)$ is zero up to $J_3/J_1 \sim 0.7$ and finite for larger values of $J_1/J_3$. The region of short--range order is thus narrowed down to $0.3 \lesssim J_3/J_1 \lesssim 0.6$.

The extrapolation was obtained by fitting a polynomial of degree~$3$ in $1/N$ to values of $S(\mathbf{q})$ for various values of~$N$. The error bars in the plot indicate the error estimates for the predictions. The data points are usually fitted very well with such a polynomial - as can be seen in figure~\ref{fig:J1J3:SQ_Nscaling}. In this figure, for selected points $J_3/J_1$, the scaling of $S(\mathbf{q})$ with $N$ is plotted and the best fit by a polynomial of $3$rd degree in $1/N$ is drawn. From this fit, we obtain the predictions for the thermodynamic limit.


\subsubsection*{$J_1-J_2$ Model}

\begin{figure}
    \begin{center}
        \includegraphics[width=0.44\textwidth]{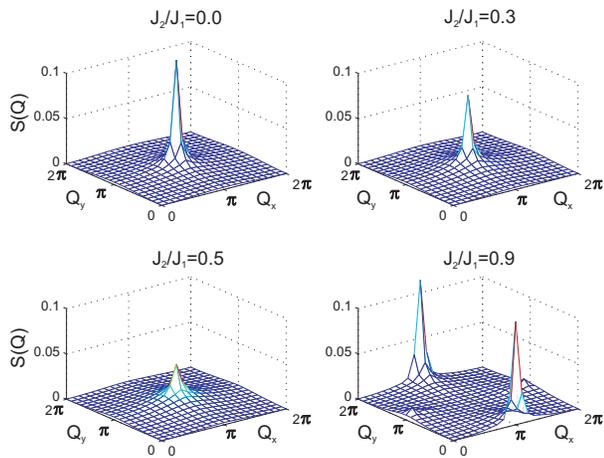}
    \end{center}
    \caption{
        Structure factor $S(\mathbf{q})$ for $J_2/J_1$ equal to $0$, $0.3$, $0.5$ and $0.9$. The results were obtained for a $14 \times 14$--lattice and $D=3$.
        }
    \label{fig:J1J2:SQ}
\end{figure}

For $J_3=0$, the static structure factor shows the following behavior: for $J_2/J_1 \lesssim 0.5$ the structure factor is peaked at $(\pi,\pi)$ which indicates long--range N\'eel order. For $J_2/J_1$ larger than $0.5$, columnar long--range order develops which is detected by a peak at $(0,\pi)$. In fact, this columnar long--range order reveals an order--by--disorder phenomenon~\cite{villain77}: quantum fluctuations select from the huge manifold of classical ground states configurations where all spins are parallel in a given direction. In the regime in--between the N\'eel and the columnar phase the peaks disappear and long--range order breaks down.
The structure factor obtained from the PEPS--calculation for $D=3$ and a $14 \times 14$ lattice is plotted in figure~\ref{fig:J1J2:SQ} for selected values of $J_2/J_1$.

\begin{figure}
    \begin{center}
        \includegraphics[width=0.44\textwidth]{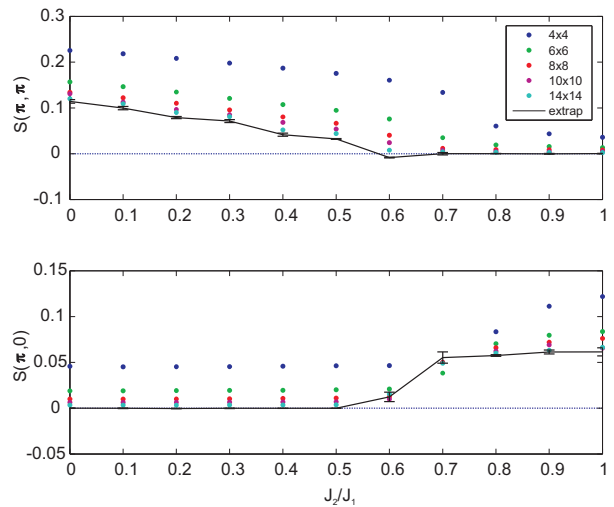}
    \end{center}
    \caption{
		Structure factors $S(\mathbf{\pi,\pi})$ (upper plot) and $S(\mathbf{\pi,0})$ (lower plot) as functions of $J_2/J_1$ for various particle--numbers~$N$. The solid lines represent extrapolations to the thermodynamic limit.
        }
    \label{fig:J1J2:SQ_pi_pi0}
\end{figure}

In figure~\ref{fig:J1J2:SQ_pi_pi0}, $S(\mathbf{\pi,\pi})$ and $S(\mathbf{\pi,0})$ are plotted as functions of $J_2/J_1$ for various particle--numbers~$N$. The extrapolation to the thermodynamic limit indicates that $S(\mathbf{\pi,\pi})$ remains finite for $J_2/J_1 \lesssim 0.5$, and that $S(\mathbf{\pi,0})$ is finite for $J_2/J_1 \gtrsim 0.5$. Thus, the systems consists of two phases with long--range order plus a possibly very small short--range order phase in the vicinity of $J_2/J_1 \sim 0.5$.


\section{Short--Range Resonating Valence Bond States} \label{sec:SRVB}

The considerable decrease of the correlation length at $J_3/J_1 \sim 0.5$ and $J_2/J_1 \sim 0.5$ opens the possibility for a short--range resonating valence bond state (SRVB) in this area. We investigate this possibility by doing a direct comparison of the PEPS results to results obtained by an exact diagonalization of the Hamiltonian in the subspace of SRVB.

\subsubsection*{$J_1-J_3$ Model}

\begin{figure}[!t]
    \begin{center}
        \includegraphics[width=0.44\textwidth]{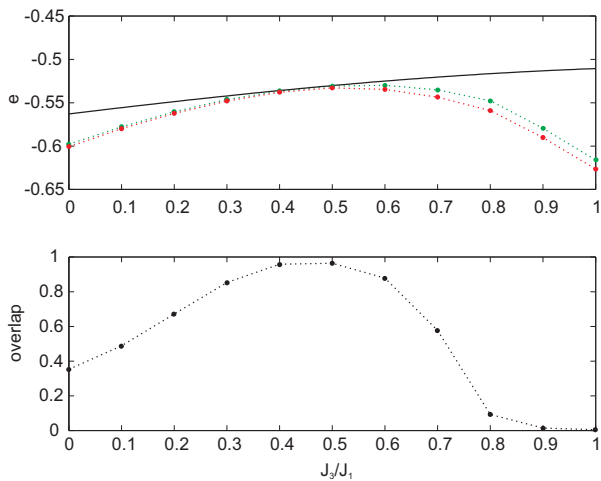}
    \end{center}
    \caption{
		\emph{upper plot:} Ground state energy of the $J_1-J_3$ model on a $6 \times 6$ lattice as a function of $J_3/J_1$, obtained by diagonalizing within the SRVB subspace (solid line) and by PEPS calculations with $D=3$ (green dots) and $D=4$ (red dots). \emph{lower plot:} Overlap between the SRVB ground state and the $D=3$--PEPS ground state.
        }
    \label{fig:J1J3:srvb}
\end{figure}

In case of the $J_1-J_3$ model, the overlap between the SRVB and the PEPS with virtual dimensions $D=3$ on a $6 \times 6$--lattice can be gathered from figure~\ref{fig:J1J3:srvb} (lower plot). As it can be seen, the overlap increases up to $99$\% at the point $J_3/J_1 = 0.5$ and is significantly smaller in other regions. A comparison of energies, however, reveals that the set of valence bond states does not cover all terms in the ground state. As shown in the upper plot of figure~\ref{fig:J1J3:srvb}, the energies of the diagonalization within the subspace of SRVB are - though very close to the PEPS--results at $J_3/J_1 \sim 0.5$ - always higher than the energies obtained within the set of PEPS. Thus, the true ground state at $J_3/J_1 \sim 0.5$ might contain a small fraction of valence bond terms that have longer range.

Even if it is assumed that the ground state is a pure SRVB in the vicinity of $J_3/J_1 = 0.5$, its properties can be very rich and it needs a more precise classification. On the one hand, it could be a state with broken translational symmetry - such as a columnar valence bond crystal or a plaquette state. On the other hand, an equally weighted superposition of valence bond states with restored translational symmetry known as spin--liquid is possible.

\begin{figure}
    \begin{center}
        \includegraphics[width=0.44\textwidth]{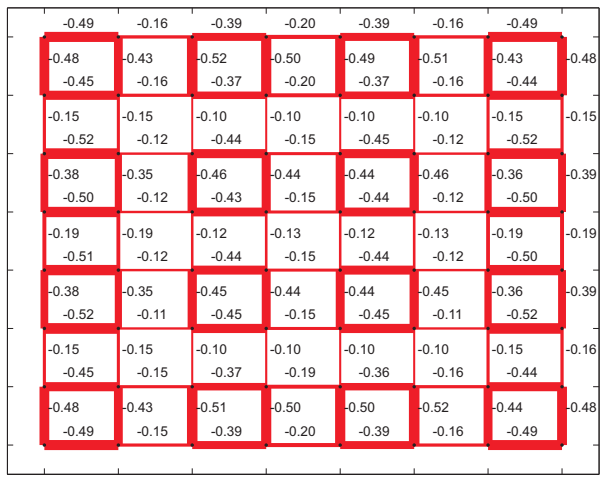}
    \end{center}
    \caption{
        Nearest--neighbor spin--spin correlations at $J_3/J_1=0.5$, calculated with $D=3$--PEPS.
        }
    \label{fig:J1J3:nncorr}
\end{figure}

The observation of nearest--neighbor spin--spin correlations $\expect{\mathbf{s}_i \cdot \mathbf{s}_j}$ gives us an indication of a plaquette state~\cite{fouet03}. These correlations are shown in figure~\ref{fig:J1J3:nncorr} for a $8 \times 8$ lattice and $D=3$. In case of a pure plaquette state, the nearest--neighbor spin--spin correlations would be equal to~$-1/2$ on a plaquette and~$0$ between two plaquettes. In our case, the values of the spin--spin correlations deviate slightly from these values, nontheless a clear plaquette structure remains visible.

The state we observe has obviously a broken translational symmetry. The plaquettes are formed between sites $(i,j)$, $(i+1,j)$, $(i,j+1)$, $(i+1,j+1)$ with $i$ and $j$ always being odd. The reason for this symmetry breaking are the chosen open boundary conditions and the even number of sites in $x$- and $y$-direction. The system chooses the configuration with a maximal number of plaquettes. In case of open boundary conditions and an even number of sites in each direction, this corresponds to the configuration of plaquettes with $i$ and $j$ being odd.

\subsubsection*{$J_1-J_2$ Model}

\begin{figure}
    \begin{center}
        \includegraphics[width=0.44\textwidth]{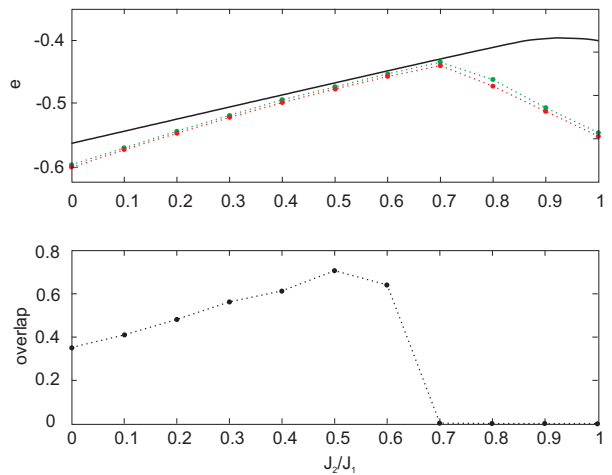}
    \end{center}
    \caption{
		\emph{upper plot:} Ground state energy of the $J_1-J_2$ model on a $6 \times 6$ lattice as a function of $J_2/J_1$, obtained by diagonalizing within the SRVB subspace (solid line) and by PEPS calculations with $D=3$ (green dots) and $D=4$ (red dots). \emph{lower plot:} Overlap between the SRVB ground state and the $D=3$--PEPS ground state.
        }
    \label{fig:J1J2:srvb}
\end{figure}

\begin{figure}
    \vspace{0.7cm}
    \begin{center}
        \includegraphics[width=0.44\textwidth]{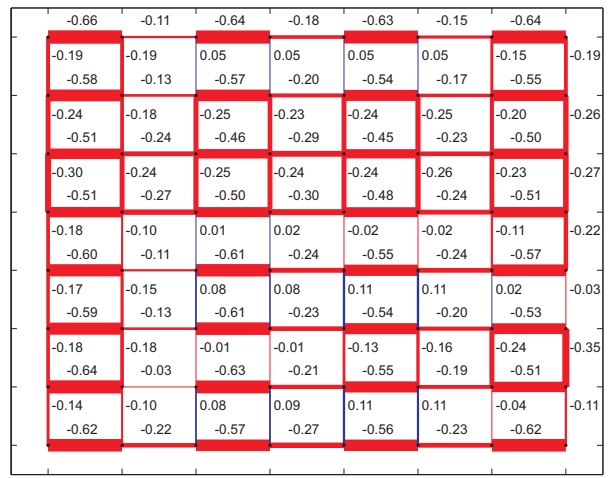}
    \end{center}
    \caption{
        Nearest--neighbor spin--spin correlations at $J_2/J_1=0.6$, calculated with $D=3$--PEPS.
        }
    \label{fig:J1J2:nncorr}
\end{figure}

The vicinity of the ground state to the class of SRVB in the region $J_2/J_1 \sim 0.5$ is analyzed in figure~\ref{fig:J1J2:srvb}. The lower plot shows the overlap of the PEPS ground state with the ground state obtained by exact diagonalization within the PEPS subspace. Lattice size $6 \times 6$ and virtual dimensions $D=3$ were considered. The overlap clearly reaches the maximum at $J_2/J_1 = 0.5$ and assumes a value of about~$70$\%. A comparison of the energies, as shown in the upper plot of figure~\ref{fig:J1J2:srvb}, uncovers that the true ground state will not be exactly in the subspace of SRVB: the energies of the PEPS calculations are slightly lower than the ones obtained from SRVB, even at the critical point. Nonetheless, the distance to the subspace of SRVB might be very small for
$J_2/J_1 \sim 0.5$.

A more precise classification of the ground state turns out to be more difficult than in the case of the $J_1-J_3$ model: observations of the nearest--neighbor spin--spin correlations $\expect{\mathbf{s}_i \cdot \mathbf{s}_j}$ yield faint indications for a columnar dimer state~\cite{leung96}. The magnitudes of the nearest--neighbor spin--spin correlations for a $8 \times 8$ lattice, $D=3$ and $J_2/J_1=0.6$ are shown in figure~\ref{fig:J1J2:nncorr}. In a pure dimer state, the spin--spin correlations are equal to~$-3/4$ on a dimer bond and zero between two dimer bonds. Even though these values are not attained, a columnar order of dimers is visible.


\begin{figure}[t]
    \begin{center}
        \includegraphics[width=0.44\textwidth]{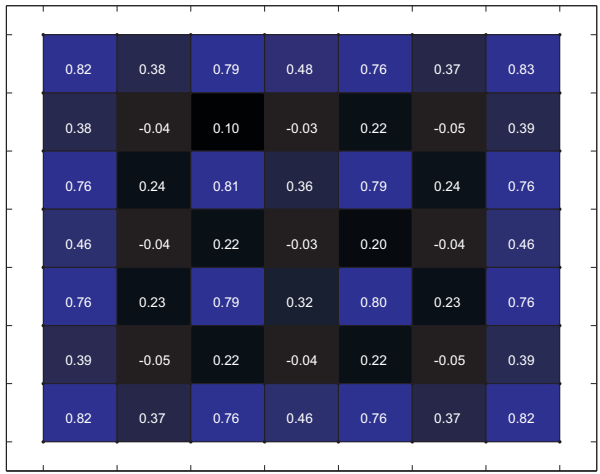}
    \end{center}
    \caption{
        Plaquette order parameter $Q_{\alpha \beta \gamma \delta}$ at $J_3/J_1=0.5$, evaluated on a $8 \times 8$ lattice. The used virtual dimension is $D=3$.
        }
    \label{fig:J1J3:plaquette1}
\end{figure}

\section{Other Order Parameters} \label{sec:other}

In order to shed more light on the properties of the ground state in the maximally frustrated regime, more elaborate order parameters are investigated like the plaquette order parameter~\cite{fouet03}, the columnar order parameter or the VBC order parameter~\cite{mambrini06}.

The plaquette order parameter~\cite{fouet03} distinguishes clearly a N\'eel ordered phase from a plaquette valence bond crystal. It is defined, using the cyclic permutation operator $P_{\alpha \beta \gamma \delta}$ of the four spins $\alpha$, $\beta$, $\gamma$ and $\delta$ on one plaquette, as
\begin{eqnarray*}
Q_{\alpha \beta \gamma \delta} & = & \frac{1}{2} \left( P_{\alpha \beta \gamma \delta} + P_{\alpha \beta \gamma \delta}^{-1} \right) \\
& = & 2 \left( \Sprod{\alpha}{\beta} \Sprod{\gamma}{\delta} + \Sprod{\alpha}{\delta} \Sprod{\beta}{\gamma} - \Sprod{\alpha}{\gamma} \Sprod{\beta}{\delta} \right) \\
& & + 1/2 \left( \Sprod{\alpha}{\beta} + \Sprod{\gamma}{\delta} + \Sprod{\alpha}{\delta} + \Sprod{\beta}{\gamma} \right) \\
& & + 1/2 \left( \Sprod{\alpha}{\gamma} + \Sprod{\beta}{\delta} + 1/4 \right) .
\end{eqnarray*}
In case of a pure plaquette state, this order parameter assumes the value~$1$ on each plaquette; between the plaquettes, its expectation value is~$1/8$. The order parameter vanishes in case of the N\'eel state which lacks the cyclic permutation symmetry.


\begin{figure}[t]
    \begin{center}
        \includegraphics[width=0.44\textwidth]{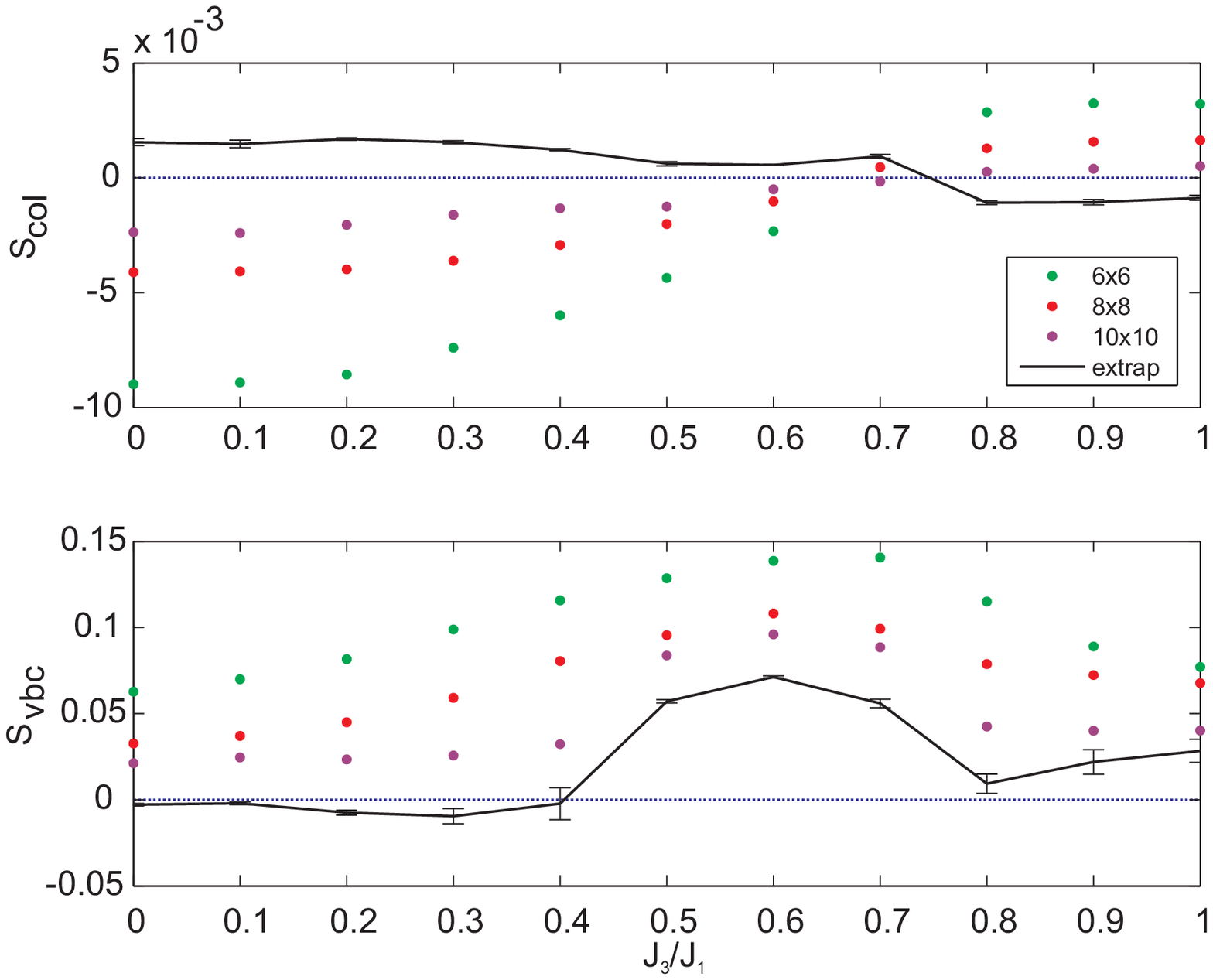}
    \end{center}
    \caption{
        Structure factors $S_{col}$ and $S_{VBC}$ as functions of $J_3/J_1$ for lattice sizes $6 \times 6$, $8 \times 8$ and $10 \times 10$, calculated using PEPS with $D=3$. The solid lines represent extrapolations to the thermodynamic limit.
        }
    \label{fig:J1J3:Scol_Svbc}
\end{figure}

\begin{figure}
    \begin{center}
        \includegraphics[width=0.44\textwidth]{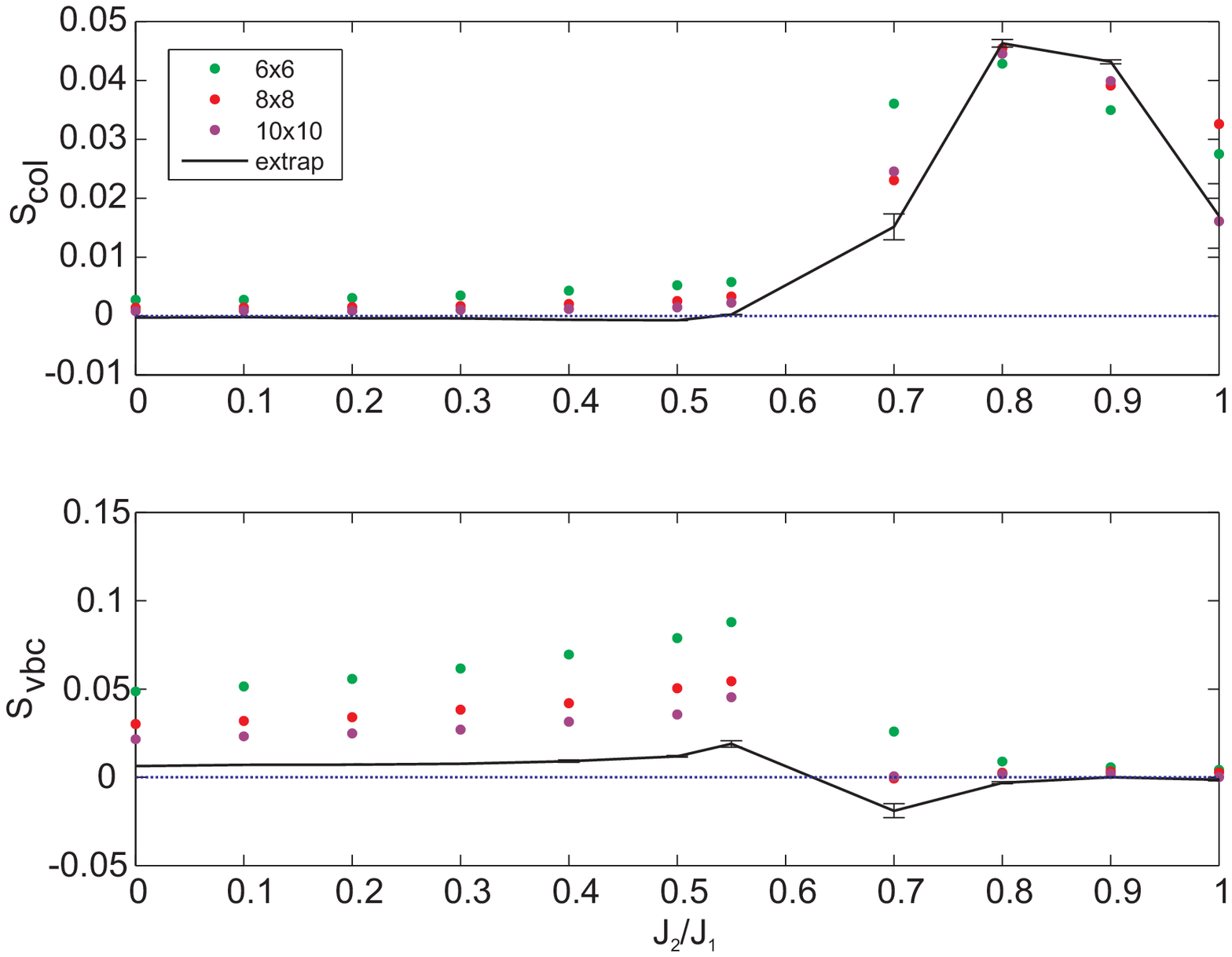}
    \end{center}
    \caption{
		Structure factors $S_{col}$ and $S_{VBC}$ as functions of $J_2/J_1$ for lattice sizes $6 \times 6$, $8 \times 8$ and $10 \times 10$, calculated using PEPS with $D=3$. The solid lines represent extrapolations to the thermodynamic limit.
        }
    \label{fig:J1J2:Scol_Svbc}
\end{figure}

The columnar order parameter and the VBC order parameter~\cite{mambrini06} are defined as structure factors of dimer--dimer correlations:
\begin{displaymath}
    S_{\lambda} = \frac{1}{N_B} \sum_{(k,l)} \varepsilon_{\lambda}(k,l)
    \left(
    \expect{(\mathbf{s}_i \cdot \mathbf{s}_j) (\mathbf{s}_k \cdot \mathbf{s}_l)}
    - \expect{(\mathbf{s}_i \cdot \mathbf{s}_j)}^2
    \right)
\end{displaymath}
Here, $k$ and $l$ run over all pairs of neighboring sites. $i$ and $j$ have fixed values and they point to two neighboring sites at the center of the lattice. $\lambda$ represents either "col" or "VBC" and the phase factor $\varepsilon_{\lambda}(k,l)$ is defined according to figure~7 in Ref.~\onlinecite{mambrini06}. $N_B$ denotes the number of terms with $\varepsilon_{\lambda}(k,l) \neq 0$.
As discussed in Ref.~\onlinecite{mambrini06}, $S_{VBC}$ and $S_{col}$ show a different behavior in the plaquette and the columnar dimer phase: in the plaquette phase, $S_{VBC}$ remains finite in the thermodynamic limit and $S_{col}$ tends to zero, whereas in the columnar dimer phase, both $S_{VBC}$ and $S_{col}$ remain finite.

All these order parameters are based on dimer--dimer correlations which are difficult to obtain with high precision from PEPS--calculations. Nonetheless, our results give a qualitative picture of the ground state.

\subsubsection*{$J_1-J_3$ Model}

In figure~\ref{fig:J1J3:plaquette1}, we plot the plaquette order parameter evaluated on all square clusters for $J_3/J_1=0.5$ and a $8 \times 8$ lattice. The results are obtained using a PEPS--Ansatz with $D=3$. A clear plaquette structure is visible, though the absolute values of the expectation values differ slightly from the optimal ones.

In figure~\ref{fig:J1J3:Scol_Svbc}, both structure factors $S_{VBC}$ and $S_{col}$ are plotted as functions of $J_3/J_1$ for various lattice sizes. The values were obtained from a PEPS-simulation with $D=3$. The solid lines indicate extrapolations to the thermodynamic limit ($N \to \infty$). The extrapolations were obtained by fitting polynomials of $1$st degree in $1/N$. It can be seen that $S_{VBC}$ is clearly peaked in the region $0.4 \lesssim J_1/J_3 \lesssim 0.8$. In addition, $S_{col}$ is $3$ orders of magnitude smaller than $S_{VBC}$ in this region. This gives us an indication for a plaquette state in the region $0.4 \leq J_1/J_3 \leq 0.8$.

Thus, the PEPS algorithm reproduces well the properties of the $J_1-J_3$ model in the regimes of weak frustration and gives strong indications for a plaquette ordered state in the regime of strong frustration. In this regime, our results are consistent with those of Ref.~\onlinecite{mambrini06}.

\subsubsection*{$J_1-J_2$ Model}

In figure~\ref{fig:J1J2:Scol_Svbc}, the structure factors $S_{VBC}$ and $S_{col}$ as functions of $J_2/J_1$ are plotted for various lattice sizes. The virtual dimension used for the PEPS-simulation was $D=3$. Again, the solid lines represent extrapolations to the thermodynamic limit. As can be seen, both quantities feature regions in which they are finite and regions in which they tend to zero rapidly. In the case of $S_{col}$, the structure factor is very small for $J_2/J_1 \lesssim 0.5$ and finite otherwise. On the other hand, $S_{VBC}$ is finite for $J_2/J_1 \lesssim 0.7$ and close to zero otherwise. This indicates the existence of a columnar dimer phase within the
region $0.5 \lesssim J_2/J_1 \lesssim 0.7$.


\vspace{1cm}

\section{Conclusions}

In conclusion, we have applied the PEPS algorithm to the $J_1-J_2$ and the $J_1-J_3$ model and discussed our observations. In both models we observed a separation in long--range order and short--range order regions. In the short--range order regions, the ground state seems to lie within the subspace of SRVB--states. Due to lack of precision we cannot settle the question of whether the ground state is a VBC or a spin--liquid, but there are strong indications that the ground state is a plaquette state in case of the $J_1-J_3$ model and a columnar dimer state in case of the $J_1-J_2$ model. Simulations with higher~$D$ and the inclusion of symmetries might lead to a more concrete answer.

We thank M. Troyer, S. Sachdev and X. Wen for discussions. Work supported by the DFG-SFB 631, EU-SCALA and the DFG excellence cluster Munich Advanced Photonics.


\end{document}